\documentclass[preprint,showpacs,aps]{revtex4}

\usepackage{graphicx}
\usepackage{amsmath}
\usepackage{amssymb}

\newcommand{\rr}{{\vec{\mathbf{r}}}}  
\newcommand{\pp}{{\vec{\mathbf{p}}}}  

\begin{document}

\title{Generalized Harmonic Oscillator and the Schr\"{o}dinger Equation
with Position-Dependent Mass}

\author{JU Guo-Xing$^{1}$\footnote{jugx@nju.edu.cn}, CAI Chang-Ying$^{1,2}$ and REN Zhong-Zhou$^{1}$ }

\affiliation{$^{1}$Department of Physics, Nanjing University,
Nanjing 210093, P. R. China\\
$^{2}$Department of Physics, Jinggangshan University, Jian 343009,
China}

\begin{abstract}
We study the generalized harmonic oscillator which  has both the position-dependent mass and
the potential depending on the form of mass function in a more general framework. The explicit expressions of the
eigenvalue and eigenfunction for such system are given, they have the same forms as those for the
usual harmonic oscillator with constant mass. The coherent state and the  its properties
for the system with PDM are also discussed.
We give the corresponding effective potentials
for several mass functions, the systems with such potentials
are isospectral to the usual harmonic oscillator.

\end{abstract}


\maketitle

\section{Introduction}

Recently, the study of quantum systems with position-dependent
mass (PDM) has attracted a lot of interests
\cite{Bas,von,mor,Ser, Bar,Lev,dek1,dek2,pla,Mil,des,gon1,gon2,Alh,roy,koc,koc2,que1,que2,que3,que4,que5,dong,
cai,roy05a,roy05b,jiang05,sch, mus1,mus2,ju06, ju07}.
The models have played  important roles in
description of the electronic properties of semiconductors
\cite{Bas}, quantum dots \cite{Ser}, liquid crystals\cite{Bar}
and so on. In the theoretical works on the system with PDM, the main concern was  the exact solutions of
the corresponding Schr\"{o}dinger
equation or its relativistic generalizations such as Klein-Gordon equation and Dirac equation.
Using coordinate transformation method\cite{man,lev,dutt}, supersymmetric quantum
mechanics \cite{wit,Coo} and other methods, many solvable potentials of the wave equations for different
mass functions have been obtained. These potentials can be classified according to the  forms of
eigenfunctions of the  wave equations, which show some similarities between  systems with constant mass
and with PDM\cite{lev,ju07}.

It is well known that the harmonic oscillator with constant mass is a very important system
in quantum mechanics\cite{mos}.
Apart from its wide applications in problems of condensed matter, atomic, nuclear and elementary particle
physics, the harmonic oscillator also has connections with many
interesting methods analytically solving Schr\"{o}dinger equation such as factorization method, supersymmetric quantum
mechanics, algebraic method. So, it is natural to generalize the usual harmonic oscillator and related studies
to the case of position-dependent mass. There exist some works in this aspect
by using point canonical transformation and Lie algebraic approach\cite{roy05a,roy05b,jiang05}. However, these
investigations only involve the specific potentials and mass functions.
In this paper, we consider the so-called generalized
harmonic oscillator(GHO) based on the following two requirements:
the Hamiltonian for the system with PDM can be factorized in terms of
the rising operator and lowering operator which are  generalization of the  creation
and destruction operators for the usual harmonic oscillator, respectively;
the rising operator and lowering operator satisfy the same form of
commutation relation as that for the  creation
and destruction operators. With these  requirements, we put the discussion of GHO in a more general framework.
We will see that many systems with PDM rather than with specific mass functions
have the same energy spectrum and same form of eigenfunctions as those of the usual
harmonic oscillator.

The paper is organized as follows. In section II, the condition for the Hamiltonian with PDM
to be that for GHO is derived, the eigenproblem of GHO is solved using the operator method.
Some  studies such as the coherent states related to the harmonic oscillator  are generalized
to the system with PDM and their properties are discussed.
In section III, for some mass functions, the corresponding potentials are given.
In the last section,
we will make some remarks.

\section{Generalized Harmonic Oscillator and its Solution}
\label{gho}

When the mass of a particle depends on its position,  the mass
and momentum operators no longer commute\cite{von}, so there are
several ways to define the kinetic energy operator. We start with the kinetic energy introduced by
von Roos\cite{von}, the Hamiltonian with position-dependent  mass $M(\rr)$ and potential
energy $V(\rr)$  reads
\begin{equation}\label{ham-vr}
\begin{aligned}
\hat{H}&=\frac{1}{4}\left[M(\rr)^\alpha\hat{\pp} M(\rr)^\beta\hat{\pp} M(\rr)^\gamma
+M(\rr)^\gamma\hat{\pp} M(\rr)^\beta\hat{\pp} M(\rr)^\alpha\right]+V(\rr)\\
&=-\frac{\hbar^{2}}{4m_{0}}\left[m(\rr)^\alpha\vec{\nabla}m(\rr)^\beta\vec{\nabla}m(\rr)^\gamma
+m(\rr)^\gamma\vec{\nabla}m(\rr)^\beta\vec{\nabla}m(\rr)^\alpha\right]+V(\rr),
\end{aligned}
\end{equation}
where $m_{0}$ is a constant mass and $m(\rr)$ is a
dimensionless position-dependent mass $M(\rr)=m_0m(\rr)$, $\alpha, \beta,\gamma$ are parameters and satisfy
the condition of $\alpha+\beta+\gamma=-1$.
 Using natural units
$(m_{0}=\hbar=1)$ and only considering one-dimensional system, we can rewrite
Hamiltonian (\ref{ham-vr}) as
\begin{equation}\label{ham-1}
\hat{H}=-\frac{1}{2}\frac{d}{dx}\frac{1}{m}\frac{d}{dx}+V_{eff}(x)
=-\frac{1}{2m}\frac{d^{2}}{dx^{2}}+\frac{m'}{2m^{2}}\frac{d}{dx}+V_{eff}(x),
\end{equation}
where the effective potential is
\begin{equation}\label{effp}
   V_{eff}(x)=V(x)
+\frac{1}{4}(\beta+1)\frac{m''(x)}{m^2}-\frac{1}{2}[\alpha(\alpha+\beta+1)+\beta+1]\frac{m'(x)^2}{m(x)^3},
\end{equation}
and $m'(x)=\frac{dm}{dx},\,m''(x)=\frac{d^2m}{dx^2}$.

There are many debates for the choice of the parameters $\alpha, \beta, \gamma$. Morrow and Brownstein\cite{mor}
have shown that $\alpha=\gamma$ based on the comparison between the experimental results and the analytical solutions of some models.
In the following, we use this kind of condition for the parameters as a basic fact.
We will introduce the  lowering and rising operators $\hat{A}$ and $\hat{A}^{+}$ so that the Hamiltonian
(\ref{ham-1}) can be factorized in terms of these operators and its solution  be obtained by using the operator method.
For this purpose, we define the  lowering and rising operators $\hat{A}$ and $\hat{A}^{+}$ as follows\cite{koc2}, respectively
\begin{equation}\label{rop}
\hat{A}=\frac{1}{\sqrt{2}}\left[m(x)^{\frac{\beta}{2}}\frac{d}{dx}m(x)^{-\frac{1}{2}(\beta+1)}+\mu(x)\right],
\end{equation}
\begin{equation}\label{lop}
\hat{A}^{+}=\frac{1}{\sqrt{2}}\left[-m(x)^{-\frac{1}{2}(\beta+1)}\frac{d}{dx}m(x)^{\frac{\beta}{2}}+\mu(x)\right],
\end{equation}
where the function $\mu(x)$ will be determined by the requirement that
\begin{equation}\label{bcr}
[\hat{A},\hat{A}^{+}]=1.
\end{equation}
The constrain (\ref{bcr}) makes our discussion be restricted to GHO, so $\mu(x)$ in Eqs.(\ref{rop}) and
(\ref{lop}) is not the so-called superpotential, which is different from that in \cite{koc2,roy05a,roy05b}.

Using Eqs.(\ref{rop}) and (\ref{lop}) and performing some calculations, we have
\begin{equation}\label{bcr2}
[\hat{A},\hat{A}^{+}]=\frac{\mu'(x)}{\sqrt{m(x)}}-\frac{1}{4m(x)}(2\beta+1)
\left[\frac{m''(x)}{m(x)}-\frac{3}{2}\left(\frac{m'(x)}{m(x)}\right)^2\right].
\end{equation}
It follows  from Eqs.(\ref{bcr}) and (\ref{bcr2}) that $\mu(x)$ and $\beta$  are given by the
following relations, respectively
\begin{equation}\label{int}
\beta=-\frac{1}{2},\;\;\;
\mu(x)=\int^x\sqrt{m(x)}dx.
\end{equation}
Now, Eqs.(\ref{rop}) and (\ref{lop}) can be rewritten as follows,
\begin{equation}\label{rlop}
\begin{aligned}
&\hat{A}=\frac{1}{\sqrt{2}}\left[m(x)^{-\frac{1}{4}}\frac{d}{dx}m(x)^{-\frac{1}{4}}+\mu(x)\right],\\
&\hat{A}^{+}=\frac{1}{\sqrt{2}}\left[-m(x)^{-\frac{1}{4}}\frac{d}{dx}m(x)^{-\frac{1}{4}}+\mu(x)\right].
\end{aligned}
\end{equation}
It is easy to see that the operators $\hat{A}$ and $\hat{A}^{+}$ will become
the  destruction and creation operators for the harmonic oscillator respectively when the mass $m(x)$ is a constant.

If we assume that the potential function $V(x)$ in Eq.(\ref{ham-1})  depends
on  mass $m(x)$  and is given by
\begin{equation}\label{pot}
V(x)=\frac{1}{2}[\mu(x)]^{2},
\end{equation}
then the Hamiltonian (\ref{ham-1})  can be rewritten as
\begin{equation}\label{ham-2}
H=\hat{A}^{+}\hat{A}+\frac{1}{2}=\hat{N}+\frac{1}{2},
\end{equation}
where $\hat{N}=\hat{A}^{+}\hat{A}$ may be called number-like operator because it reduces to the
number operator for the harmonic oscillator when the mass $m(x)$ is
independent of  $x$. It is also evident that with
$m(x)=1$ the potential (\ref{pot}) and the Hamiltonian
(\ref{ham-2}) become  the potential
and  Hamiltonian for the harmonic oscillator, respectively. In this sense, we call the above system with PDM a
generalized harmonic oscillator.

With Eq.(\ref{bcr}), we can easily get the commutators  between operators $\hat{A}$, $\hat{A}^{+}$ and $\hat{N}$
\begin{equation}\label{bcr3}
[\hat{N},\hat{A}]=-\hat{A},\;\;\;\;[\hat{N},\hat{A}^{+}]=\hat{A}^{+}.
\end{equation}
The following procedure solving the eigenproblem of Eq.(\ref{ham-2}) is similar to that of the
harmonic oscillator\cite{lan}.
If we assume that $\psi_{n}(x)$ is an eigenstate of $\hat{N}$ with
eigenvalue $n$, that is
\begin{equation}\label{nef}
\hat{N}\psi_{n}(x)=n\psi_{n}(x),
\end{equation}
then using Eq.(\ref{bcr3}), we have the relation
\begin{equation}\label{cre}
\hat{N}\hat{A}^{+}\psi_{n}(x)=(n+1)\hat{A}^{+}\psi_{n}(x).
\end{equation}
Eq.(\ref{cre}) indicates that $\hat{A}^{+}\psi_{n}(x)$ is also an eigenstate of
$\hat{N}$ with eigenvalue $(n+1)$, so $\hat{A}^{+}$ can be
regarded as a raising operator. By a similar reasoning, $\hat{A}$ is a
lowering operator which means that $\hat{A}\psi_{n}(x)$ is an
eigenstate of $\hat{N}$ with eigenvalue $(n-1)$. If we assume the eigenstate of $\hat{N}$
has a lower bound, that is there exists a state $\psi_{0}(x)$ satisfying the relation
\begin{equation}\label{grs}
\hat{A}\psi_{0}(x)=\frac{1}{\sqrt{2}}\left[m(x)^{-\frac{1}{4}}\frac{d}{dx}(m(x)^{-\frac{1}{4}}\psi_{0}(x))
+\mu(x)\psi_{0}(x)\right]
=0,
\end{equation}
then all eigenstates of $\hat{N}$ can be obtained by successive
application of the operator $\hat{A}^{+}$ on the state
$\psi_{0}(x)$. The existence of the lower bound is related to the requirement
that $|\hat{A}\psi_{n}(x)|^2$ is real and positive.
Solving Eq.(\ref{grs}), we get the analytical expression of $\psi_{0}(x)$
\begin{equation}\label{gsf}
\psi_{0}(x)=[m(x)]^{\frac{1}{4}}e^{-\frac{1}{2}[\mu(x)]^{2}}.
\end{equation}
Now, acting $\psi_{0}(x)$ on the left with the operator $\hat{A}^{+}$, we have
\begin{equation}\label{fe1}
\psi_{1}(x)=\hat{A}^{+}\psi_{0}(x)=\sqrt{2}\mu(x)\psi_{0}(x)
=\frac{\sqrt{2}}{2}[m(x)]^{\frac{1}{4}}e^{-\frac{1}{2}[\mu(x)]^{2}}H_1(\mu(x)).
\end{equation}
Similarly , we get
\begin{equation}\label{fe2}
\psi_{2}(x)=(\hat{A}^{+})^2\psi_{0}(x)=\{2[\mu(x)]^2-1\}\psi_{0}(x)
=\frac{1}{2}[m(x)]^{\frac{1}{4}}e^{-\frac{1}{2}[\mu(x)]^{2}}H_2(\mu(x)).
\end{equation}
By induction, one can show
that all eigenstates of $\hat{H}$ are given by
\begin{equation}\label{hef}
\psi_{n}(x)={\cal N}_n[m(x)]^{\frac{1}{4}}e^{-\frac{1}{2}[\mu(x)]^{2}}H_{n}(\mu(x)),\
\ \ (n=0,1,2,...),
\end{equation}
where ${\cal N}_n$ is the normalization coefficient, $\mu(x)$ is defined by Eq.(\ref{int}), and $H_{n}(\mu(x))$
is the Hermite polynomial. The normalization of $\psi_n$ means that
\begin{equation}\label{wnor}
\begin{aligned}
 \int_{-\infty}^{\infty}|\psi_{n}(x)|^2dx=&1={\cal N}_n^2
 \int_{-\infty}^{\infty}[m(x)]^{\frac{1}{2}}e^{-[\mu(x)]^{2}}H_{n}^2(\mu(x))\,dx \\
 =&{\cal N}_n^2
 \int_{\mu_{min}}^{\mu_{max}}e^{-[\mu(x)]^{2}}H_{n}^2(\mu(x))\,d\mu(x),
 \end{aligned}
\end{equation}
where $\mu_{min}=\mu(-\infty)$ and $\mu_{max}=\mu(\infty)$. If $\mu_{min}=0$ or $-\infty$ and $\mu_{max}=\infty$, we can get
the explicit expression of ${\cal N}_n$ as follows
\begin{equation}\label{norc}
    {\cal N}_n=
\left\{
\begin{aligned}
&\frac{1}{\sqrt{2^nn!\sqrt{\pi}}},\;\;\;\;(\mu_{min}=-\infty)\\
&\frac{1}{\sqrt{2^{n-1}n!\sqrt{\pi}}}.\;\;\;\;(\mu_{min}=0)
\end{aligned}
\right.
\end{equation}
The requirement for the parameters $\mu_{min}, \mu_{max}$ will input constraint on the mass functions. However,
if we require that $\psi_{n}(x)$ is
orthogonal for different $n$, then there is only one choice of $\mu_{min}=-\infty$.

Using Eqs. (\ref{ham-2}) and (\ref{nef}), we have
\begin{equation}
\hat{H}\psi_{n}(x)=\left(n+\frac{1}{2}\right)\psi_{n}(x),\ \ \ \ (n=0,1,2,...),
\end{equation}
which means that $\psi_{n}(x)$ is the  eigenstate  of $\hat{H}$
with eigenvalue $E_n=n+\frac{1}{2}$.
In a sense, GHO is in the same class as the harmonic oscillator considering that they have the
same energy spectrum but with different
potentials.

Now, we can calculate the matrix elements $\langle n|\hat{A}^{+}|n'\rangle$,  $\langle n|\hat{A}|n'\rangle$ which
are the same  as those for the harmonic oscillator
\begin{equation}\label{me}
\begin{aligned}
&\langle n|\hat{A}^{+}|n'\rangle=\int_{-\infty}^{\infty} \psi_n \hat{A}^{+} \psi_{n'}\, dx
=\sqrt{n'+1}\int_{-\infty}^{\infty} \psi_n  \psi_{n'+1}\, dx=\sqrt{n'+1} \delta_{n,n'+1},\\
&\langle n|\hat{A}|n'\rangle=\int_{-\infty}^{\infty} \psi_n \hat{A} \psi_{n'}\, dx
=\sqrt{n'}\int_{-\infty}^{\infty} \psi_n  \psi_{n'-1}\, dx=\sqrt{n'} \delta_{n,n'-1}.
\end{aligned}
\end{equation}

Because the operators $\hat{A}$ and $\hat{A}^{+}$ have the same kind of properties of
the creation and destruction operators, many studies on the latter may be generalized to the former.
For example, the canonical coherent state is defined to be the eigenstate of the destruction operator\cite{gla}.
If we generalize this definition of the coherent state to the operator $\hat{A}$, that is,
\begin{equation}\label{cs}
  \hat{A}|z\rangle= z|z\rangle,
\end{equation}
where $z$ is a complex number, and  similarly we expand $|z\rangle$ in term of the state $|n\rangle$,
$|z\rangle=\sum\limits_{n=0}^{\infty}c_n|n\rangle$, then we have
\begin{equation}\label{cs2}
    |z\rangle=e^{-\frac{|z|^2}{2}}\sum\limits_{n=0}^{\infty}\frac{z^n
    }{\sqrt{n!}}|n\rangle,
\end{equation}
where the factor $e^{-\frac{|z|^2}{2}}$ comes from the convention of $\langle z|z\rangle=1$.
The completeness relations, non-orthogonality for different eigenstates $z$ and $z'$, and other properties
for the coherent state of system with constant mass  are also hold for the system with PDM.
However, due to the fact
that we cannot express $x$  in terms of the
combinations of the operators $\hat{A}$ and $\hat{A}^{+}$ as in the case of harmonic oscillator, it is difficult to
calculate the matrix elements such as $\langle z|x|z\rangle,\;\langle z|x^2|z\rangle$. From Eqs.(\ref{rlop}), we have
\begin{equation}
\begin{aligned}
&\mu(x)=\frac{1}{\sqrt{2}}(\hat{A}+\hat{A}^{+}),\\
&\hat{\pi}(x)=m^{-\frac{1}{4}}\hat{p}m^{-\frac{1}{4}}
=-i m^{-\frac{1}{4}}\frac{d}{dx}m^{-\frac{1}{4}}=-\frac{i}{\sqrt{2}}(\hat{A}-\hat{A}^{+}),
\end{aligned}
\end{equation}
where $\hat{\pi}$ is the so-called deformed momentum\cite{que4}. With Eq.(\ref{cs}), it is
easy to get the following expressions of the matrix elements
\begin{equation}\label{mecs}
\begin{aligned}
  & \langle z|\mu(x)|z\rangle= \frac{1}{\sqrt{2}}(z+z^{*}),
  &&\langle z|\mu(x)^2|z\rangle= \frac{1}{2}[(z+z^{*})^2+1],\\
   & \langle z|\hat{\pi}(x)|z\rangle=- \frac{i}{\sqrt{2}}(z-z^{*}),
   &&\langle z|\hat{\pi}(x)^2|z\rangle= -\frac{1}{2}[(z-z^{*})^2-1],
\end{aligned}
\end{equation}
and
\begin{equation}\label{meancs}
\begin{aligned}
&\Delta \mu(x)=\sqrt{\langle z|\mu(x)^2|z\rangle-(\langle z|\mu(x)|z\rangle)^2}=\frac{1}{\sqrt{2}},\\
&\Delta \hat{\pi}(x)=\sqrt{\langle z|\hat{\pi}(x)^2|z\rangle-(\langle z|\hat{\pi}(x)|z\rangle)^2}=\frac{1}{\sqrt{2}}.
\end{aligned}
\end{equation}
Noticing that $\mu(x)$ and $\hat{\pi}(x)$ are  functions of $x$ except that $\hat{\pi}(x)$ is related to the
momentum $\hat{p}$, Eqs.(\ref{meancs})
show that there exist more quantities with minimum uncertainty in the coherent states for systems with
PDM than for those with constant mass.

It should be emphasized that above discussion is the generalization
of the method used for the usual harmonic oscillator. In this
setting, the analytical expressions of the eigenvalues and
eigenfunctions for GHO are obtained through the so-called Heisenberg
algebra generated by $\hat{A}^{+},\hat{A},\hat{N}$ and the unit
operator $1$ rather than the Lie algebra $su(1,1)$ or  the nonlinear
algebra studied in \cite{roy05a,roy05b}. It is also noted that the
condition (6) was used to determine the superpotential in the
framework of supersymmetry quantum mechanics in
\cite{roy05a,roy05b}. However, the condition (6) has several
functions in our discussions: (i) determining the parameters in the
Hamiltonian (\ref{ham-1}); (ii) indicating that the potential in the
Hamiltonian (\ref{ham-1}) is related to the mass function when the
Hamiltonian is required to be factorized with the operators obeying
the condition (6); (iii) showing that the algebraic methods and some
related studies to the usual harmonic oscillator may be applicable
to the systems with PDM. As to (iii), the coherent state for the
system of PDM discussed above is such an example and it gives a more
general result on minimum uncertainty than that for the case of
constant mass. In addition, we also study the orthonormality of the
wavefunctions and its possible restriction to the form of the mass
functions, which is illustrated through examples in the next
section. To our knowledge, all those mentioned above are not
appeared in the literature for the systems with PDM.

\section{Mass Functions and Potentials}

In the last section, we have shown that the generalized harmonic oscillators have the same energy spectrums and
the same form eigenstates as those  of the harmonic oscillators for a lot of mass functions.
In this section, we will give the corresponding potentials for several mass functions that
satisfy the requirements for the orthonormality of  $\psi_{n}(x)$.

{\bf Example\;1.}  We first consider the mass
function $m(x)$  to be of the form
\begin{equation}\label{efm-1}
m(x)=\biggl(\frac{a+x^{2}}{1+x^{2}}\biggr)^{2},
\end{equation}
where $a>0$. This effective mass was used by many authors in their studies on  solvability of the systems with PDM
\cite{pla,Alh,roy,koc2,roy05a,roy05b}. It is noted that we have $m(x)=1$ for $a=1$, which corresponds to the harmonic oscillator.

Substituting Eq.(\ref{efm-1}) into Eq.(\ref{int}), we have
\begin{equation}
\mu(x)=x+(a-1)\arctan x,
\end{equation}
and the related potential function  given by
\begin{equation}
V_{eff}(x)=\frac{1}{2}[\mu(x)]^{2}+\frac{(a-1)}{2}\frac{[3x^{4}+(4-2a)x^{2}-a]}{(a+x^{2})^{4}},
\end{equation}
which is the effective-mass analogue of the singular oscillator potential.
From Eqs.(\ref{hef}) and (\ref{efm-1}), we get the eigenstate as follows
\begin{equation}
\psi_{n}(x)={\cal N}_n\sqrt{\frac{a+x^{2}}{1+x^{2}}}e^{-\frac{1}{2}[\mu(x)]^{2}}H_{n}(\mu(x)),\
\ \ (n=0,1,2,...).
\end{equation}
These results are the same as those obtained by  the coordinates transformation method, SUSY
quantum mechanics method and so on\cite{pla,Alh,roy, roy05a}.

{\bf Example\;2.} The mass for  a particle is exponentially
increasing or decreasing  which has the form\cite{des,gon1,gon2}
\begin{equation}\label{efm-2}
m(x)=e^{a x}.
\end{equation}
This kind of mass function may be used in the study of confine energy states for carriers in
semiconductor quantum well structures\cite{gon1,gon2}. For the mass function (\ref{efm-2}),
Eq.(\ref{int}) gives
\begin{equation}
\mu(x)=\left\{
\begin{aligned}
&\frac{2}{a}e^{\frac{a}{2}x},  && \;\;\;\; (a\neq0), \\
&x, & &\;\;\;\;(a=0),
\end{aligned}
\right.
\end{equation}
and $a=0$ is related to $m(x)=1$, that is to the case of constant mass.
From Eq.(\ref{pot}), we have the potential
\begin{equation}\label{e2pot}
V(x)=\left\{
\begin{aligned}
&\frac{2}{a^2}e^{ax},  && \;\;\;\; (a\neq0), \\
&\frac{1}{2}x^2, & &\;\;\;\;(a=0),
\end{aligned}
\right.
\end{equation}
which indicates it has a similar behavior as the mass (\ref{efm-2}) for $a\not=0$\cite{des}.
The  effective potential corresponding to the
mass (\ref{efm-2}) can be written as
\begin{equation}
V_{eff}(x)=\frac{1}{2}[\mu(x)]^{2}-\frac{3a^{2}}{32}e^{-ax}.
\end{equation}
The corresponding wavefunctions are
\begin{equation}
\psi_{n}(x)={\cal N}_ne^{\frac{a}{4}x}e^{-\frac{1}{2}[\mu(x)]^{2}}H_{n}(\mu(x)),\ \ \ (n=0,1,2,...).
\end{equation}
It is noted that the orthogonality of the eigenstate  requires $a<0$ for $x<0$ and $a>0$ for $x>0$.

{\bf Example\;3.} We consider another  mass which reads as
\begin{equation}\label{efm-3}
m(x)=1+\tanh(ax),
\end{equation}
where $a$ is a real parameter. The solution of Schr\"odinger equation with mass (\ref{efm-3}) and some smooth potential
was studied in Refs.\cite{dek1} and \cite{dek2} by solving Schr\"odinger equation directly. In this
case, we get the expression for $\mu(x)$ in Eq. (\ref{int})
\begin{equation}
\mu(x)=\left\{
\begin{aligned}
&\frac{\sqrt{2}}{a}\ln(e^{ax}+\sqrt{1+e^{2ax}}), & & \;\;\;\; (a\neq0), \\
&x, && \;\;\;\;(a=0).
\end{aligned}
\right.
\end{equation}
The effective potential corresponding to the mass (\ref{efm-3}) is
\begin{equation}
V_{eff}(x)=\frac{1}{2}[\mu(x)]^{2}-\frac{a^{2}}{16}\frac{4 e^{2ax}+3}{ e^{2ax}( e^{2ax}+1)},
\end{equation}
and the wavefunctions are
\begin{equation}
\psi_{n}(x)={\cal N}_n[1+\tanh(ax)]^{\frac{1}{4}}e^{-\frac{1}{2}[\mu(x)]^{2}}H_{n}(\mu(x)),\
\ \ (n=0,1,2,...).
\end{equation}

{\bf Example\;4.} The mass is a power of $x$, that is
\begin{equation}\label{efm-4}
m(x)=x^{a},
\end{equation}
where $a$ is a real parameter. For this mass, $\mu(x)$ in Eq. (\ref{int}) takes the form
\begin{equation}
    \mu(x)=
\left\{
\begin{aligned}
&\frac{2}{a+2}x^{(a+2)/2},&&\;\;(a\not=-2),\\
&\ln x, &&\;\;(a=-2).
\end{aligned}
\right.
\end{equation}
The effective potential and the wavefunctions related to the mass (\ref{efm-4}) are, respectively
\begin{equation}
    V_{eff}=\frac{1}{2}[\mu(x)]^2-\frac{1}{32}a(3a+4)x^{-(a+2)},
\end{equation}
\begin{equation}
\psi_{n}(x)={\cal N}_n x^{\frac{a}{4}}e^{-\frac{1}{2}[\mu(x)]^{2}}H_{n}(\mu(x)),\
\ \ (n=0,1,2,...).
\end{equation}

{\bf Example\;5.} The mass is a deformed hyperbolic function of $x$
\begin{equation}\label{efm-5}
m(x)={\rm sech}^2(ax),
\end{equation}
where $a$ is a positive parameter. Now, $\mu(x)$ in Eq. (\ref{int}) is
\begin{equation}\label{efp-5}
    \mu(x)=\frac{2}{a}\arctan e^{ax}.
\end{equation}
The corresponding effective potential  is
\begin{equation}
    V_{eff}=\frac{1}{2}[\mu(x)]^2-\frac{1}{16}[3\cosh(2ax)+1].
\end{equation}
The wavefunctions are
\begin{equation}\label{wf-5}
\psi_{n}(x)={\cal N}_n [{\rm sech}(ax)]^{\frac{1}{2}}e^{-\frac{1}{2}[\mu(x)]^{2}}H_{n}(\mu(x)),\
\ \ (n=0,1,2,...).
\end{equation}

{\bf Example\;6.} The mass function has the form \cite{jiang05}
\begin{equation}\label{efm-6}
m(x)=\frac{a^2}{(q+x^2)^2},
\end{equation}
where $a$ is a real parameter and $q>0$. When $q=0$, (\ref{efm-5}) reduces to that of (\ref{efm-4}).
In this case, $\mu(x)$ in Eq. (\ref{int}) is
\begin{equation}\label{efp-6}
    \mu(x)=\frac{a}{\sqrt{q}}\arctan\left(\frac{x}{\sqrt{q}}\right).
\end{equation}
The  effective potential  is
\begin{equation}
    V_{eff}= \frac{a^2}{2q}\left[\arctan\left(\frac{x}{\sqrt{q}}\right)\right]^2
    -\frac{q}{2a^2}-\frac{x^2}{a^2}.
\end{equation}
The wavefunctions are
\begin{equation}\label{wf-6}
\psi_{n}(x)={\cal N}_n \sqrt{\frac{a}{q+x^2}}e^{-\frac{1}{2}[\mu(x)]^{2}}H_{n}(\mu(x)),\
\ \ (n=0,1,2,...).
\end{equation}
It is noted that mass (\ref{efm-6}) has no analogue of the constant mass for whatever $a$
 and $q$, which is different from the mass functions in the other five examples. Also, the orthogonality
of the wavefunction for GHO requires that $\mu(x)$ have the property $\mu_{min}=-\infty$ and $\mu_{max}=\infty$, but
$\mu(x)$ in Eqs. (\ref{efp-5}) and (\ref{efp-6}) do not so. In this sense, the mass functions
(\ref{efm-5}) and (\ref{efm-6}) should be excluded
from  a possible choice of the mass, and the wavefunctions in (\ref{wf-5}) and (\ref{wf-6}) are only formal.
Note that the wavefunctions in \cite{jiang05} are unnormalized.

\section{Remarks and discussions}

In the above sections, we have discussed a special system with PDM, i.e. the generalized harmonic
oscillator that can be solved using operator method.
We obtain the analytical expressions of its energy spectrums and eigenstates. For several mass functions,
the corresponding effective potentials are given. In our discussions, we restrict the lowering and rising operators
to obey the commutation relation (\ref{bcr}) which is same as that for the creation and
destruction operators of the harmonic oscillator.
With this kind of construction, we  have restrict the choice of the parameters $\alpha, \beta, \gamma$
in the Hamiltonian (\ref{ham-1}). If we further require that the Hamiltonian (\ref{ham-1})
can be factorized into the product
of the lowering and rising operators, then the potential $V(x)$ is determined by the mass function,
which is consistent with
that from the SUSY quantum mechanics method.
Different from the SUSY method, the key of our discussions
lies in that we can generalize a lot of investigations of  the
harmonic oscillator to GHO. For example, the coherent state in section \ref{gho} for the system with PDM
is studied with this kind of consideration,
and it indeed shows some differences
from that for the system of constant mass. Also, the orthonormality of the wavefunctions
for GHO puts the restriction on the choice of the mass functions.

\section*{Acknowledgment}

The program is supported  by
the National Natural Science Foundation of China under contract
No. 10125521, No. 60371013 and by the 973 National Major State Basic Research
and Development of China under contract No. G2000077400.


\begin{thebibliography}{99}

\bibitem{Bas}G. Bastard, {\em Wave  Mechanics  Applied  to Semiconductor
Heterostructure}, 1988,Les Ulis: Editions de Physique.

\bibitem{von}O. von Roos, {\em Phy. Rev.}, {\bf B27}(1983)7547.

\bibitem{mor}R. A. Morrov and K. R. Brownstein, {\em Phy. Rev.}, {\bf B30}(1984)678.

\bibitem{Ser}L. I. Serra  and E. Lipparini,  {\em Europhys. Lett.}, {\bf 40}(1997)667.

\bibitem{Bar}M. Barranco, M. Pi, S. M. Gatica, E. S. Hernandez  and J. Navarro,
{\em Phys. Rev.}, {\bf B56}(1997)8997.

\bibitem{Lev}J. M. Levy-Leblond,  {\em Phys. Rev.}, {\bf A52}(1995)1845.

\bibitem{dek1}L. Dekar, L. Chetouani and T.  F. Hammann, {\em J. Math. Phys.}, {\bf 39}(1998)2551.

\bibitem{dek2}L. Dekar, L. Chetouani and T.  F. Hammann,  {\em Phys.  Rev.}, {\bf A59}(1999)107.

\bibitem{pla}A. R. Plastino, A. Rigo, M. Casas, F. Garcias and A. Plastino,
{\em Phys. Rev.}, {\bf A60}(1999)4318.

\bibitem{Mil}V. Milanovic and Z. Ikovic,{\em J.  Phys. A: Math. Gen.}, {\bf 32}(1999)7001.

\bibitem{des}A. de Souza Dutra  and C. A. S. Almeida, {\em Phys. Lett.}, {\bf A275}(2000)25.

\bibitem{gon1}B. G\"{o}n\"{u}l, B. G\"{o}n\"{u}l,D. Tutcu and O. \"{O}zer,
{\em Mod. Phys. Lett.}, {\bf A17}(2002)2057.

\bibitem{gon2}B. G\"{o}n\"{u}l, O. \"{O}zer, B. G\"{o}n\"{u}l  and F. \"{U}zg\"{u}n,
{\em Mod. Phys. Lett.}, {\bf A17}(2002)2453.

\bibitem{Alh}A. D. Alhaidari, {\em Phys.  Rev.}, {\bf A66}(2002)042116.

\bibitem{roy}B. Roy and P. Roy, {\em J.  Phys. A: Math. Gen.}, {\bf 35}(2002)3691.

\bibitem{koc}R. Koc  and M. Koca,  {\em J.  Phys. A: Math. Gen.}, {\bf 36}(2003)8105.

\bibitem{koc2}R. Koc and H. T\"ut\"unc\"uler, {\em  Ann. Phys. (Leipzig)},{\bf 12}(2003)684.


\bibitem{que1}C. Quesne  and V. M. Tkachuk,   {\it J.  Phys. A: Math. Gen.}, {\bf 36}(2003)10373.

\bibitem{que2}C. Quesne  and V. M. Tkachuk,  {\it J.  Phys. A: Math. Gen.}, {\bf 37}(2004)10095.

\bibitem{que3}C. Quesne  and V. M. Tkachuk,  {\it J.  Phys. A: Math. Gen.}, {\bf 37}(2004) 4267.

\bibitem{que4}B. Bagchi, A. Banerjee, C. Quesne  and V. M. Tkachuk,  {\it J.  Phys. A: Math. Gen.}, {\bf 38}(2005)2929.

\bibitem{que5}C. Quesne, {\it Ann. Phys.(N. Y.)}, {\bf 321}(2006)1221.

\bibitem{dong}S. H. Dong and M. Lozada-Cassou,   {\em Phys. Lett.}, {\bf A337}(2005)313.

\bibitem{cai} C. Y. Cai,  Z. Z. Ren and  G. X. Ju, {\it Commun. Theor. Phys.}, {\bf 43}(2005)1019.

\bibitem{roy05a} B. Roy, {\it Europhys. Lett.}, {\bf 72} (2005) 1.

\bibitem{roy05b} B. Roy, P. Roy, {\it Phys. Lett.}, {\bf A 340} (2005) 70.

\bibitem{jiang05}L. Jiang, L. Z. Yi, C.S. Jia, {\it Phys. Lett.}, {\bf A 345} (2005)279.

\bibitem{sch}A. G. M. Schmidt,   {\em Phys. Lett.}, {\bf A353}(2006)459.

\bibitem{mus1} O. Mustafa and S. H. Mazharimousavi, {\it J.  Phys. A: Math. Gen.}, {\bf 39}(2006)10537.

\bibitem{mus2} O. Mustafa and S. H. Mazharimousavi, {\it Phys. Lett.}, {\bf A358}(2006)259.

\bibitem{ju06} G. X. Ju, Y Xiang and Z. Z. Ren,  {\it Commun. Theor. Phys.}, {\bf 46}(2006)819,quant-ph/0601005.

\bibitem{ju07} G. X. Ju,  C. Y. Cai, Y Xiang and Z. Z. Ren, {\it Commun. Theor. Phys.}, {\bf
47}(2007)1001, quant-ph/0601004.


\bibitem{man}M. F. Manning,  {\em Phys. Rev.}, {\bf 48}(1935)161.

\bibitem{lev}G. Levai, {\em J. Phys. A: Math. Gen.}, {\bf 22}(1989)689.

\bibitem{dutt}R. De  Dutt and U. Sukhatme, {\em J. Phys. A: Math. Gen.}, {\bf 25}(1992)L843.

\bibitem{wit}E. Witten, {\em Nucl. Phys.}, {\bf B185}(1981)513.

\bibitem{Coo}F. Cooper, A. Khare and U. Sukhatem, {\em Phys. Rep.}, {\bf 251}(1995)267.

\bibitem{mos}M. Moshinsky and Y. F. Smirnov, {\em The Harmonic Oscillator in Modern Physics}, 1996,
Amsterdam:Harwood Academic Publishers.

\bibitem{lan}L. D. Landau  and E. M. Lifshitz, {\em Quantum
Mechanics, Non-Relativistic Theory}, 3rd ed., 1977, London: Pergamon
Press.

\bibitem{gla}R. J. Glauber, {\em Phys. Rev.}, {\bf 131}(1963)2766.


\end{thebibliography}
\end{document}